\documentclass[aps,twocolumn,floatfix,nofootinbib]{revtex4-1}
\usepackage{graphicx,amsmath,amssymb,verbatim,color}
\usepackage{booktabs}
\usepackage{comment}
\usepackage{soul}
\usepackage[dvipsnames]{xcolor}
\usepackage{bm}
\usepackage[utf8]{inputenc}
\usepackage[colorlinks=true,citecolor=blue,linkcolor=blue,urlcolor=blue, backref=false,pdfborder={0 0 0}]{hyperref}
\usepackage{float}
\usepackage{multirow}

\newcommand{\be}{\begin{equation}\begin{gathered}}
\newcommand{\ee}{\end{gathered}\end{equation}} 
\newcommand{\barr}{\begin{eqnarray}}
\newcommand{\earr}{\end{eqnarray}} 
\usepackage{romannum}
\usepackage[normalem]{ulem}

\newcommand{\bs}{\boldsymbol}
\newcommand{\ov}{\overline}

\newcommand{\bk}{\bs{k}}
\newcommand{\bv}{\bs{v}_{\rm bc}}
\newcommand{\hk}{\hat{k}}

\begin{document}

\pagenumbering{arabic}

\title{Magnetic fields from small-scale primordial perturbations}
\author{Nanoom Lee}
\email{nanoom.lee@nyu.edu}
\author{Yacine Ali-Ha\"imoud}
\email{yah2@nyu.edu}
\affiliation{Center for Cosmology and Particle Physics, Department of Physics, New York University, New York, New York 10003, USA}

\date{\today} 
\begin{abstract}

Weak magnetic fields must have existed in the early Universe, as they were sourced by the cross product of electron density and temperature gradients through the Biermann-battery mechanism. In this paper we calculate the magnetic fields generated at cosmic dawn by a variety of small-scale primordial perturbations, carefully computing the evolution of electron density and temperature fluctuations, and consistently accounting for relative velocities between baryons and dark matter. We first compute the magnetic field resulting from standard, nearly scale-invariant primordial adiabatic perturbations, making significant improvements to previous calculations. This ``standard" primordial field has a root mean square (rms) of a few times $10^{-15}$ nG at $20\lesssim z \lesssim 100$, with fluctuations on $\sim$ kpc comoving scales, and could serve as the seed of present-day magnetic fields observed in galaxies and galaxy clusters. In addition, we consider early-Universe magnetic fields as a possible probe of non-standard initial conditions of the Universe on small scales $k \sim 1\text{--}10^3$ Mpc$^{-1}$. To this end, we compute the maximally-allowed magnetic fields within current upper limits on small-scale adiabatic and isocurvature perturbations. Enhanced small-scale adiabatic fluctuations below current Cosmic Microwave Background spectral-distortion constraints could produce magnetic fields as large as $\sim 5\times 10^{-11}$ nG at $z = 20$. Uncorrelated small-scale isocurvature perturbations within current Big-Bang Nucleosynthesis bounds could potentially enhance the rms magnetic field to $\sim 10^{-14}\text{--}10^{-10}$ nG at $z = 20$, depending on the specific isocurvature mode considered. While these very weak fields remain well below current observational capabilities, our work points out that magnetic fields could potentially provide an interesting window into the poorly constrained small-scale initial conditions of the Universe.
\end{abstract}
\maketitle

\section{Introduction}

While magnetic fields are ubiquitous in the Universe \cite{Vallee:2004osq,Neronov:2010gir,Beck:2012,Durrer:2013pga}, their origin remains unknown. A possible explanation could be the dynamo amplification of ``primordial" magnetic fields, generated in the early Universe prior to non-linear structure formation \cite{Durrer:2013pga}. Indeed, even seed magnetic fields as weak as $10^{-21}$ nG at the time of galaxy formation could be sufficient to source the magnetic fields observed today in galaxies and galaxy clusters \cite{Davis:1999bt}.

In 1950, Biermann showed that magnetic fields can be generated starting from vanishing initial conditions in an ionized plasma if the electron density and temperature gradients are misaligned \cite{biermann1950ursprung}, a process now known as the Biermann-battery mechanism. About a decade ago, Naoz and Narayan \cite{Naoz:2013wla} (hereafter NN13), considered this mechanism within a cosmological context for the first time. They computed the magnetic field sourced by linear fluctuations originating from adiabatic initial conditions, assuming the latter have a nearly scale-invariant power spectrum with amplitude and slope consistent with large-scale Cosmic Microwave Background (CMB) anisotropy data. They found that these ``standard" adiabatic initial conditions could lead to a magnetic field with root-mean-square (rms) $\sim 10^{-16}\text{--}10^{-15}$ nG at redshift $z \sim 10\text{--}100$, fluctuating on comoving scales of order $10$ kpc. NN13 accounted approximately for the effect of supersonic relative motions between baryons and cold dark matter (CDM) \cite{Tseliakhovich:2010bj}, which they showed to significantly affect the resulting magnetic field power spectrum.

A byproduct of NN13's work is the understanding that the cosmological Biermann-battery mechanism is most efficient on $\sim$ 1--10 kpc comoving scales. This implies that the resulting magnetic field could, in principle, offer a window into initial conditions on these scales, much smaller and much more poorly constrained than the $\sim 1\text{--}10^3$ Mpc scales directly probed by CMB-anisotropy and large-scale structure data. Not only could kpc-scale initial conditions have a significantly larger amplitude than their Mpc--Gpc scale counterparts, they may also depart significantly from adiabaticity, since isocurvature (i.e.~non-adiabatic) initial conditions are especially weakly constrained on such scales. The purpose of the present paper is to compute the magnetic field generated by non-standard initial conditions through the Biermann-battery mechanism, in the hopes that this observable may eventually serve as a probe of the small-scale primordial Universe. 

A related endeavor was recently undertaken by Flitter et al.~\cite{Flitter:2023xql} (hereafter F+23), whose work inspired ours. Specifically, F+23 computed the magnetic field sourced by small-scale compensated isocurvature perturbations (CIPs), in which initial baryon and CDM perturbations compensate each other to produce an exactly vanishing total matter fluctuation. In this work, we significantly expand on F+23's calculation in multiple aspects. First and foremost, we consider a variety of isocurvature initial conditions in addition to CIPs, as well as amplified small-scale adiabatic initial conditions. Second, we numerically evolve linear electron density and temperature fluctuations, rather than using simple analytic approximations which can be inaccurate by up to one order of magnitude. Last but not least, we incorporate exactly the effect of large-scale relative motions between baryons and CDM, which significantly affect the growth of small-scale perturbations, be they originated from adiabatic or isocurvature initial conditions. Along the way, we also make some significant improvements to NN13's original work. We update their prediction for the magnetic field sourced by standard adiabatic initial conditions, which we find can reach root-mean-square (rms) of $\sim4\text{--}5\times10^{-15}$nG at $20\lesssim z \lesssim100$, peaking at $z\sim60$, assuming small-scale initial conditions consistent with the Planck 2018 best-fit cosmology \cite{Planck2018}.  

Our main result is to predict the magnetic field produced by non-standard small-scale perturbations, and can be summarized as follows. We find that enhanced small-scale adiabatic initial conditions can lead to magnetic fields as large as $\sim 10^{-11}$ nG without violating CMB spectral-distortion constraints \cite{Chluba:2013dna}. We also show that, within current constraints, small-scale isocurvature perturbations may source magnetic fields reaching $\sim 10^{-10}$ nG at $z=20$, depending on the specific type of isocurvature mode. We note that our more accurate treatment leads to a prediction for the magnetic field sourced by CIPs up to one order of magnitude larger than that of F+23. Although to our knowledge no observational method currently exists to probe such weak small-scale magnetic fields at cosmic dawn (but see Ref.~\cite{Venumadhav:2014tqa,Gluscevic:2016gns} for a probe of large-scale magnetic fields of comparable magnitude), these field amplitudes are in principle large enough to seed magnetic fields observed in today's galaxies and galaxy clusters, and can be significantly larger than the expectation from standard adiabatic initial conditions. 

The rest of this paper is organized as follows. In Sec.~\ref{sec:evolution}, we describe how we calculate perturbations in electron density and baryonic gas temperature, which are the principal ingredients of the Biermann-battery mechanism, including the effect of relative velocity between CDM and baryons. In Sec.~\ref{sec:biermann-battery} we derive the power spectrum of the magnetic field sourced by general electron and temperature fluctuations through the Biermann-battery mechanism, and present our results with standard primordial adiabatic perturbations. In Sec.~\ref{sec:PB-probe} we consider non-standard adiabatic and isocurvature initial conditions saturating current constraints, and compute the resulting magnetic field power spectrum. We conclude in Sec.~\ref{sec:discussion}. We give a pedagogical derivation of the Biermann-battery mechanism in a cosmological context in Appendix \ref{app:Biermann}, provide details of our numerical implementation in Appendix \ref{app:numerics}, and give a detailed comparison of our work with that of F+23 in Appendix \ref{appendix:PB-cip}.

\section{Evolution of Perturbations}
\label{sec:evolution}

\subsection{Implementation}
\label{subsec:evolution}

In this section we describe how we compute the linear evolution of perturbations in CDM, baryons, free electron fraction $x_e$, and gas temperature $T_{\rm gas}$, the latter two being relevant to the computation of magnetic fields. We separate perturbations into ``large scales" and ``small scales", depending on whether their comoving wavenumber $k$ is less or greater than 1 Mpc$^{-1}$, respectively. 

On large scales, perturbations are insensitive to relative velocities between baryons and CDM (their effect is only relevant at $k \gtrsim 40$ Mpc$^{-1}$ \cite{Tseliakhovich:2010bj}). Moreover, baryon pressure is negligible, and as a consequence free-electron and gas temperature fluctuations do not affect the evolution of large-scale perturbations at linear level (even though the converse is true). We may therefore obtain the evolution of baryon density and velocity perturbations $(\delta_b, \theta_b)$ from the Boltzmann code \textsc{class} \cite{class}, and compute the linear response of the free-electron fraction and gas temperature to these perturbations using the modified recombination code \textsc{hyrec-2} \cite{Ali-Haimoud:2010tlj,Ali-Haimoud:2010hou,Lee:2020obi}, with the method described in Ref.~\cite{Lee:2021bmn}. This allows us to extract the relative perturbations $\delta_{x_e} \equiv \delta x_e/x_e$ and $\delta_{T_{\rm gas}} \equiv \delta T_{\rm gas}/T_{\rm gas}$ as a function of time (or redshift) and wavenumber, for any given initial conditions.

The same is approximately true for the evolution of small-scale perturbations prior to kinematic decoupling of baryons from photons, at $z \approx 1060$, since at that epoch the dynamics of baryons is dominated by their coupling to photons. Therefore, we also use \textsc{class} combined with the modified \textsc{hyrec-2} linear-response solution for $\delta_{x_e}, \delta_{T_{\rm gas}}$ for small-scale perturbations\footnote{While perturbed recombination calculations \cite{Naoz:2005pd,Lewis:2007zh,Senatore:2008vi} are available in \textsc{class}, in its default setting the code works only for relatively large-scale modes. Moreover, the perturbed recombination equation implemented in \textsc{class} are only valid in the limit when photoionizations are negligible, and after transitions from the excited to the ground state are no longer bottlenecked, i.e.~for $z \lesssim 900$. Except for our neglect of photon perturbations in the gas temperature evolution, our implementation with the modified \textsc{hyrec-2} is therefore more accurate.} until $z = 1060$. After photon-baryon kinematic decoupling, baryon pressure becomes relevant, and as consequence the evolution of baryon fluid variables becomes coupled to gas temperature fluctuations, which are themselves coupled to free-electron fraction fluctuations. Moreover, relative velocities between baryons and CDM affect the growth of small-scale perturbations, as they advect baryon and CDM perturbations relative to one another \cite{Tseliakhovich:2010bj}. Since the effect of baryon-CDM relative velocities is non-linear and not included into \textsc{class}, we implement our own differential equation solver, initialized with our \textsc{class} + modified \textsc{hyrec-2} solution at $z = 1060$. We follow the treatment of Refs.~\cite{Tseliakhovich:2010bj, Ali-Haimoud:2013hpa}. Given that scales $k \geq 1$ Mpc$^{-1}$ are well within the horizon scale at $z \leq 1060$, we may solve the fluid equations in the subhorizon limit. We may also neglect photon perturbations, due to their rapid free-streaming. Moreover, given that relative velocities fluctuate on scales $k \sim 0.01-0.1$ Mpc$^{-1}$, and only affect perturbations on scales $k \gtrsim 40$ Mpc$^{-1}$, we may use moving-background perturbation theory \cite{Tseliakhovich:2010bj}, by solving for the evolution of small-scale perturbations on a local patch of uniform relative velocity $\bv \propto 1/a$, in the baryon's rest-frame. Explicitly, we solve the following system of coupled differential equations for the baryon and CDM density perturbations ($\delta_b, \delta_c$) and velocity divergences with respect to proper space ($\theta_b, \theta_c)$, as well gas temperature and ionization fraction perturbations $(\delta_{T_{\rm gas}}, \delta_{x_e})$, where overdots denote derivatives with respect to coordinate time:
\barr
\dot{\delta}_c &=& \frac{i}{a}(\bm{v}_{bc}\cdot\bm{k})\delta_c - \theta_c,\nonumber\\
\dot{\theta}_c &=& \frac{i}{a}(\bm{v}_{bc}\cdot\bm{k})\theta_c- 2H\theta_c  - \frac{3H^2}{2}(\Omega_c\delta_c+\Omega_b\delta_b),\nonumber\\
\dot{\delta}_b &=& -\theta_b,\nonumber\\
\dot{\theta}_b &=& - 2H\theta_b - \frac{3H^2}{2}(\Omega_c\delta_c+\Omega_b\delta_b) + \frac{c_s^2k^2}{a^2}\big(\delta_b+\delta_{T_{\rm gas}}\big) \nonumber\\
&&- \frac{4\rho_\gamma}{3\rho_b} n_e\sigma_T\theta_b, \nonumber\\
\dot{\delta}_{T_\text{gas}} &=& -\frac{2}{3}\theta_b + \Gamma_c\left[\frac{T_\gamma - T_{\rm gas}}{T_{\rm gas}}\delta_{x_e} - \frac{T_\gamma}{T_{\rm gas}}\delta_{T_{\rm gas}}\right],\nonumber\\
\dot{\delta}_{x_e} &=& \frac{\dot{x}_e}{x_e}\Bigg[\delta_{x_e} + \delta_b + \frac{\partial \ln \mathcal{A}_B}{\partial \ln T_{\rm gas}} \delta_{T_{\rm gas}}\nonumber\\
&&~~~~~+ \frac{\partial \ln C}{\partial \ln R_{\text{Ly}\alpha}}\left(\frac{\theta_b}{3H} - \delta_b\right)  \Bigg]. 
\label{eq:evolution}
\earr
In these equations, all quantities except for the perturbation variables $(\delta_b, \delta_c, \theta_b, \theta_c, \delta_{T_{\rm gas}}, \delta_{x_e})$ are by default background (i.e.~homogeneous) quantities, $H$ is the Hubble rate, and $\Omega_{b, c}$ are the contributions of baryons and CDM to the total critical energy density. We calculate the background gas temperature $T_{\rm gas}$ and free electron fraction $x_e$ using the recombination code \textsc{hyrec-2} \cite{Ali-Haimoud:2010tlj,Ali-Haimoud:2010hou,Lee:2020obi} implemented in \textsc{class} \cite{class}. The last term in the baryon momentum equation, proportional to the background electron density $n_e$ and the Thomson scattering cross section $\sigma_T$, accounts for residual photon drag after $z = 1060$, and is required to obtain accurate results, especially for non-adiabatic initial conditions. The baryon's isothermal sound speed squared is given by $c_s^2 \equiv T_{\rm gas}/\mu m_{\rm H}$, where $\mu\equiv [1+(m_{\rm He}/m_{\rm H})x_{\rm He}]/(1+x_{\rm He}+x_e)$ is the mean molecular weight which depends on the background free-electron fraction $x_e$ and the constant ratio of helium to hydrogen by number $x_{\rm He}$. Note that we do not include fluctuations of the molecular weight since $x_e \ll 1$ and free-electron fraction fluctuations remain very small at the redshifts of interest \cite{Ali-Haimoud:2013hpa}. The evolution of gas temperature perturbations is governed by the Compton heating rate
\be
\Gamma_c \equiv \frac{8\sigma_Ta_rT_\gamma^4}{3(1+x_{\rm He}+x_e)m_e}x_e,
\ee
and the evolution of ionization perturbations depends on the effective case-B recombination coefficient $\mathcal{A}_B(T_\gamma, T_{\rm gas})$ \cite{Ali-Haimoud:2010tlj}, as well as on the corresponding effective photoionization rate $\mathcal{B}_B(T_\gamma)$ through Peebles' $C$-factor \cite{Peebles:1968ja}
\be
C \equiv \frac{3R_{\text{Ly}\alpha} + \Lambda_{2s,1s}}{3R_{\text{Ly}\alpha} + \Lambda_{2s,1s} + 4\mathcal{B}_B},
\ee
where $\Lambda_{2s,1s}$ is the two-photon decay rate, and $R_{\text{Ly}\alpha}$ is the Lyman-$\alpha$ net decay rate defined as
\be
R_{\text{Ly}\alpha} \equiv \frac{8\pi H}{3\lambda_{\text{Ly}\alpha}^3(1-x_e)n_{\rm H}},
\ee
where $n_{\rm H}$ is the number density of hydrogen. 

Note that the equation for ionization fraction perturbations is only valid for scales larger than the typical distance travelled by Lyman-$\alpha$ photons between emission and absorption events, i.e.~for scales $k \lesssim 10^3$ Mpc$^{-1}$ \cite{Venumadhav_2015}. We will extrapolate our results to scales $k \leq 10^4$ Mpc$^{-1}$, but the reader should keep in mind that they can only be fully trusted for $k \lesssim 10^3$ Mpc$^{-1}$. 

\subsection{Initial conditions and results}

We consider five types of primordial perturbations (defined at $z \gg 10^3$, prior to horizon entry): adiabatic (AD), baryon isocurvature (BI), CDM isocurvature (CI), joint baryon and CDM isocurvature (BCI), and compensated isocurvature perturbations (CIPs). For BCI, baryons and CDM perturbations are initially equal, while for CIPs, $\delta_c = -(\Omega_b/\Omega_c)\delta_b$ initially, such that baryons and CDM produce no net gravitational potential. Let us note that, in principle, all primordial perturbations may contribute to large-scale relative velocities. However, we checked that the dominant contribution is always coming from adiabatic perturbations under current constraints on the amplitude of large-scale primordial perturbations. Hence, regardless of the type of small-scale initial conditions considered when solving Eqs.~\eqref{eq:evolution}, the relative velocity of CDM and baryon is always assumed to be determined by that of adiabatic perturbations.

Each type of initial condition $i$ is characterized by a single scalar variable $\Delta^i(\bk)$, which is the primordial curvature perturbation for adiabatic initial conditions, the initial baryon density for BI, BCI and CIP, and the initial CDM density for CI. The output of our code are the transfer functions $\mathcal{T}_\alpha^i$ of each variable $\delta_\alpha$, for each type $i$ of initial conditions, defined through 
\be
\delta_\alpha^i(z, \bk, \bv^{\rm ini}) = \mathcal{T}_\alpha^i(z, k, \bv^{\rm ini} \cdot \hk) \Delta^i(\bk), \label{eq:transfer}
\ee
where $\bv^{\rm ini}$ (which we will simply denote by $\bv$ in what follows) is the local baryon-CDM relative velocity at $z_{\rm ini} = 1060$, which affects perturbations of wave-vector $\bk$ only through its projection along $\hk$. As a cross check of our code, we have explicitly computed the matter power spectrum for adiabatic initial conditions, averaged it over the Gaussian distribution of relative velocities, and confirmed it agrees with the results of Ref.~\cite{Tseliakhovich:2010bj}.

\begin{figure*}[hb!]
\centering
\includegraphics[width=0.49\linewidth, trim= 10 0 5 20]{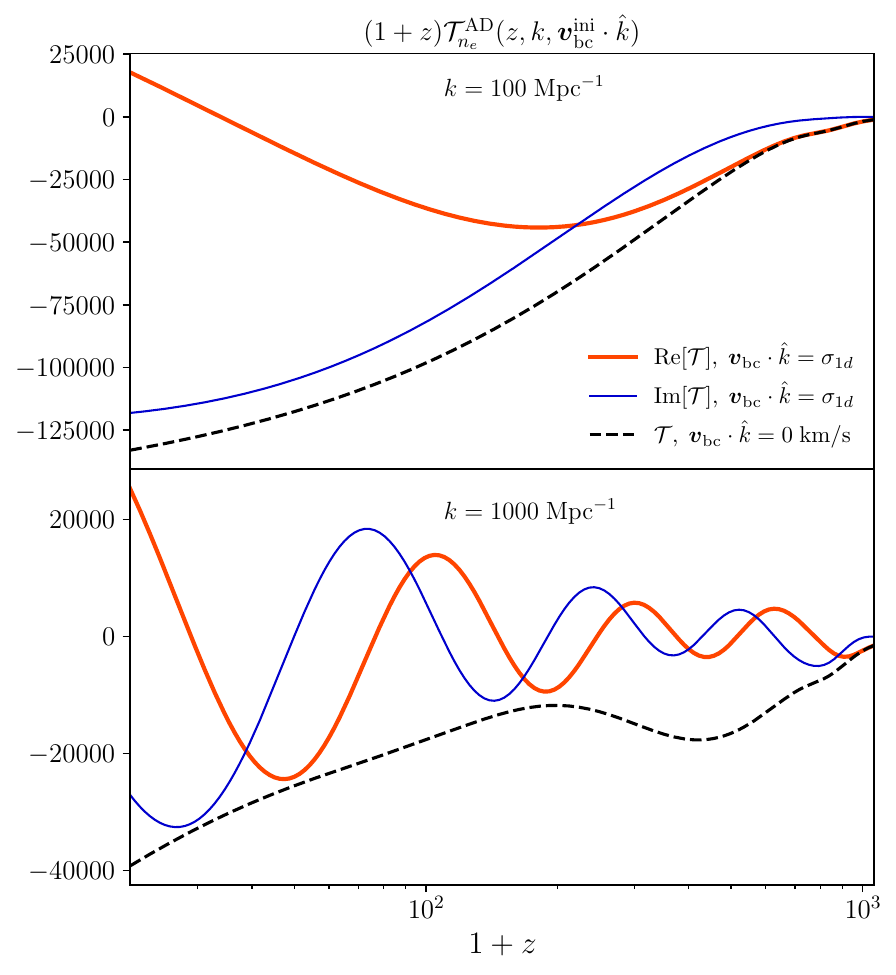}
\includegraphics[width=.482\linewidth, trim= 5 0 10 20]{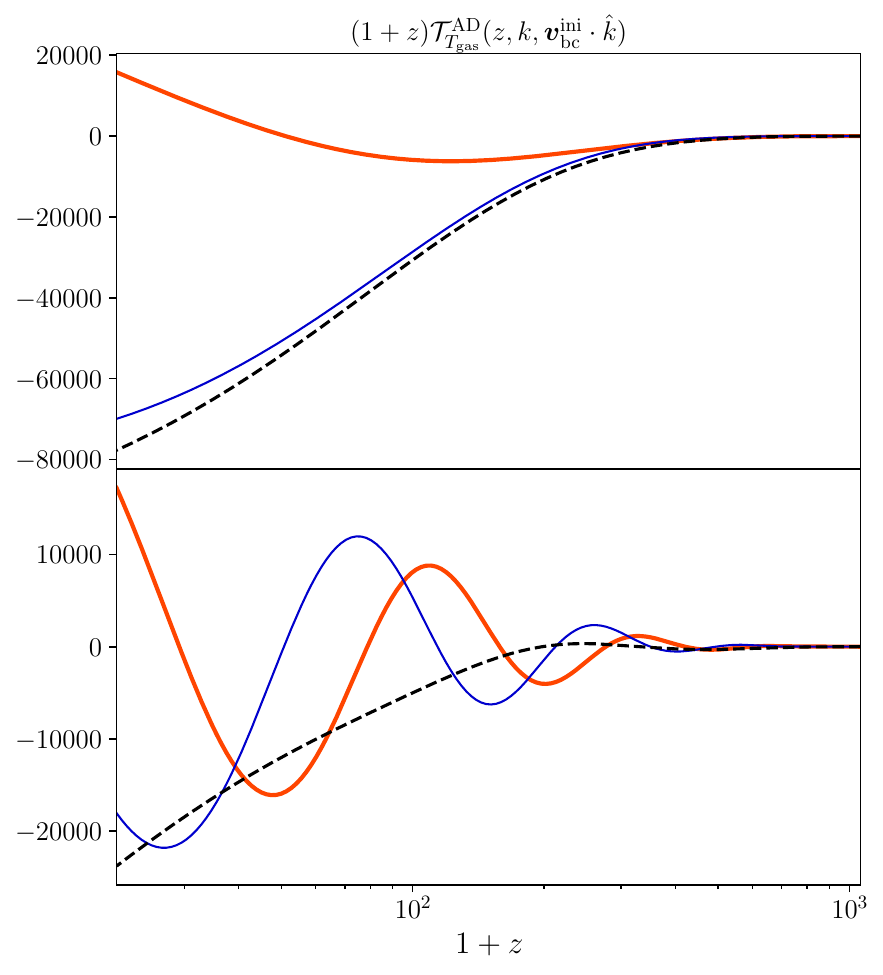}
\includegraphics[width=0.475\linewidth, trim= 0 15 5 10]{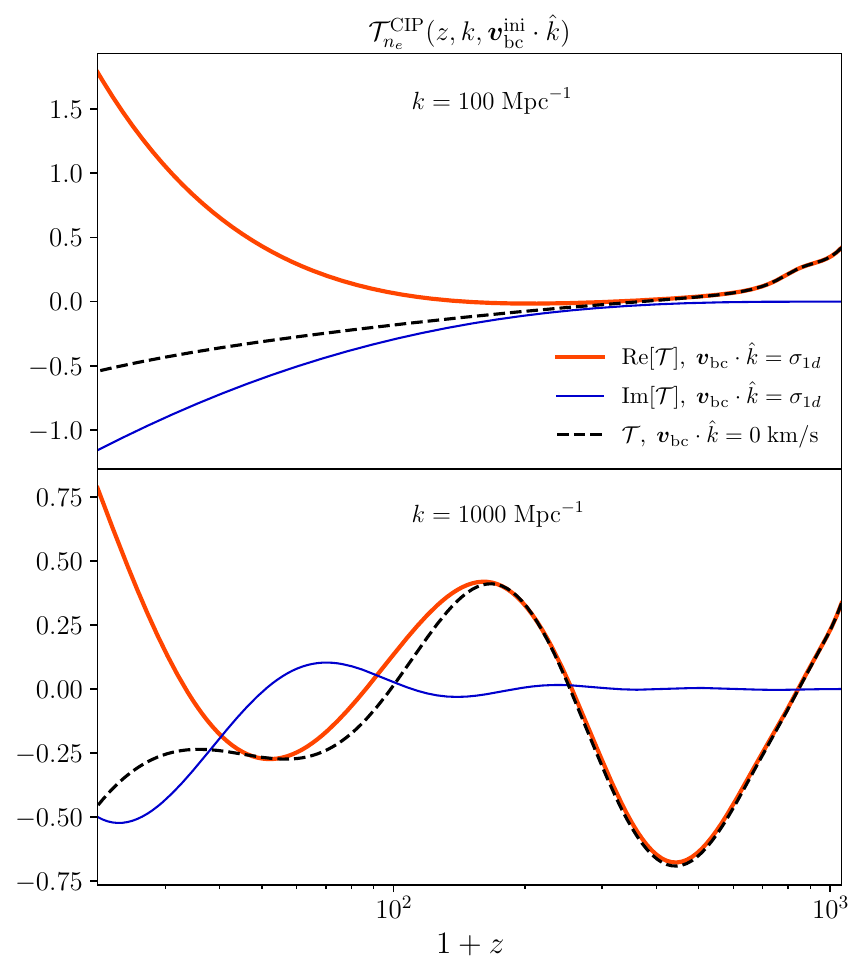}
\includegraphics[width=.48\linewidth, trim= -10 19 15 10]{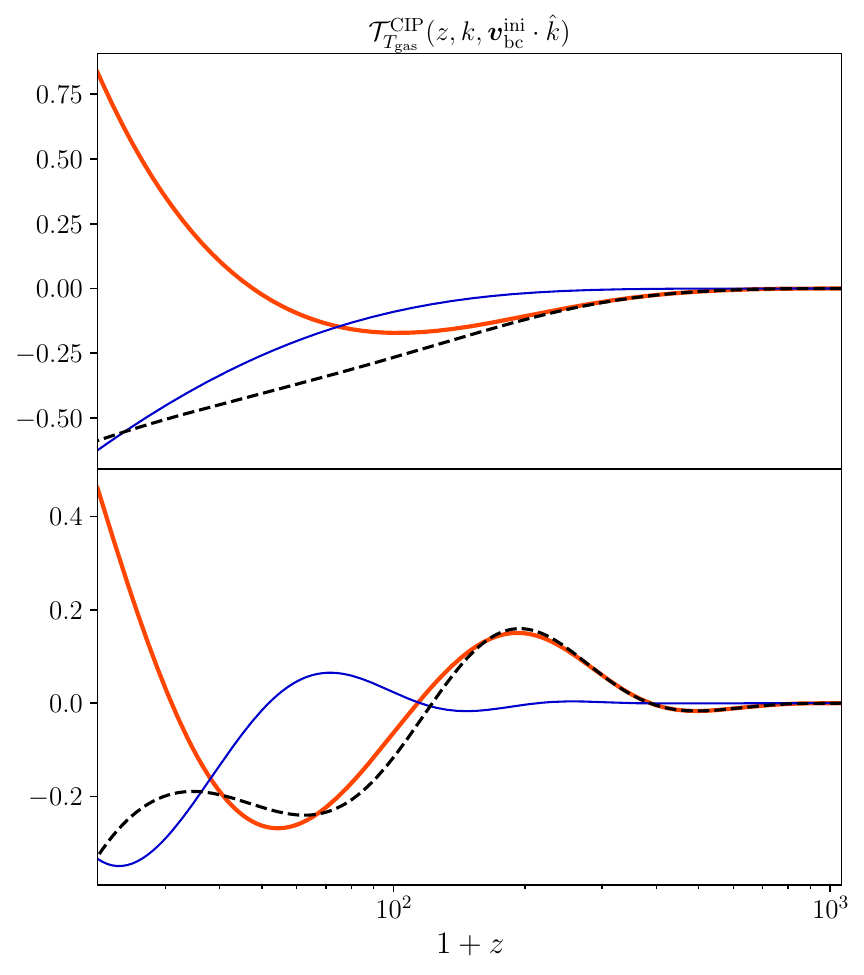}
\caption{Transfer functions for $\delta_{n_e}$ (left column) and $\delta_{T_{\rm gas}}$ (right column), for adiabatic initial conditions (top panels) and CIPs (bottom panels). The upper (lower) plot of each panel is for $k=100\;\text{Mpc}^{-1}$ ($k=1000\;\text{Mpc}^{-1}$). For $\bv\cdot\hat{k}=0$, transfer functions are real, shown in black dashed. For $\bv\cdot\hat{k} \neq 0$, transfer functions are complex functions, hence we show both their real (in red) and imaginary (in blue) parts for $\bv\cdot\hat{k} = \sigma_{1d}$, where $\sigma_{1d} \approx 17\;\text{km/s}$ is the rms relative velocity per axis.}
\label{fig:transfer}
\end{figure*}

In the context of magnetic fields, we are especially interested in the transfer functions of the gas temperature perturbation $\delta_{T_{\rm gas}}$ and of the free-electron density perturbation $\delta_{n_e}  = \delta_{x_e} + \delta_b$, since these are the relevant fields for the Biermann-battery mechanism. We show some of these transfer functions in Fig.~\ref{fig:transfer} for AD and CIP modes.

\section{Magnetic field from the Biermann-battery mechanism}
\label{sec:B-field}
\label{sec:biermann-battery}

\subsection{General expressions}

The Biermann-battery mechanism sources a magnetic field at a rate proportional to the cross product of the free-electron density and temperature gradients. In the expanding Universe, this rate is \cite{Naoz:2013wla,Papanikolaou:2023nkx,Flitter:2023xql}
\be
\frac{d}{dt}\big(a^2\bm{B}\big) = -\frac{T_{\rm gas}}{e}\bm{\nabla} \delta_{n_e} \times \bm{\nabla}\delta_{T_{\rm gas}},
\label{eq:dBdt-real}
\ee
where $e$ is the electron charge (see Appendix \ref{app:Biermann} for a pedagogical derivation). We may easily integrate this equation, and write its explicit solution in Fourier space, for a given initial value of the local relative velocity $\bv \equiv \bv(z_{\rm ini} = 1060)$:
\be
\bm{B}(t, \bm{k},\bv) = \frac{1}{[a(t)]^2e} \int_{0}^t dt'\; T_{\rm gas}(t') \\
\times \int \frac{d^3k_1}{(2\pi)^3}[\bm{k}_1\times \bm{k}]
 \delta_{n_e}(t', \bm{k}_1,\bm{v}_{bc}) \delta_{T_{\rm gas}}(t', \bm{k}-\bm{k}_1,\bm{v}_{bc}).~~~~
 \label{eq:B}
\ee
Recalling the definition of the transfer functions in Eq.~\eqref{eq:transfer}, and accounting for a possible superposition of multiple types of initial conditions, we get
\barr
\bm{B}(t, \bm{k},\bv) = \frac{1}{[a(t)]^2e} \int \frac{d^3k_1}{(2\pi)^3}[\bm{k}_1\times \bm{k}] \nonumber\\
\times \sum_{i, j} X_{eT}^{ij}(t, \bk_1, \bk - \bk_1, \bv) \Delta^i(\bk_1) \Delta^j(\bk - \bk_1),
\earr
where $i, j = $ AD, BI, CI, BCI, CIP, and we have defined
\barr
&&X_{eT}^{ij}(t, \bk_1, \bk_2, \bv) \equiv \int_{0}^t dt' ~T_{\rm gas}(t') \nonumber\\
&&~~~~~~~ \times \mathcal{T}_{n_e}^i(t', k_1, \bv \cdot \hk_1) \mathcal{T}_{T_{\rm gas}}^j(t', k_2, \bv \cdot \hk_2).\label{eq:X}
\earr
We see that, for a given realization of the initial conditions, $\bs{B}(\bk)$ depends on the full vector $\bv$, in contrast with the linear fields $\delta_{n_e}(\bk)$, which only depend on $\bv \cdot \hk$.

We now compute the small-scale power spectrum $P_B(t, \bk, \bv)$ of the magnetic field, for a given local value $\bv$. It is defined as 
\be
\langle \bs{B}(\bk) \cdot \bs{B}^*(\bk') \rangle = (2 \pi)^3 \delta_{\rm D}(\bs{k'} - \bk) P_B(\bk),
\ee
under the assumption of Gaussian initial conditions, uncorrelated between different types:
\be
\langle \Delta^i(\bk) \Delta^{j *}(\bk') \rangle =  \delta_{ij} (2 \pi)^3 \delta_{\rm D}(\bs{k'} - \bk) P^i(k).
\ee
Note that our calculation could easily be generalized to correlated initial conditions. Using the properties of Gaussian random fields, we arrive at 
\barr
P_B(t, \bk, \bv) = \frac12 \frac{1}{a^4 e^2} \int \frac{d^3 k_1}{(2 \pi)^3} |\bk_1 \times \bk|^2~~~~~~~~~ \nonumber\\
\times \sum_{i, j} P^i(k_1) P^j(|\bk - \bk_1|) \mathcal{F}_{eT}^{ij}(t,\bk_1, \bk- \bk_1, \bv), \label{eq:PB}
\earr
where we have defined 
\be
\mathcal{F}_{eT}^{ij}(t, \bk_1, \bk_2, \bv) \equiv \\
\left | X_{eT}^{ij}(t, \bk_1, \bk_2, \bv) - X_{eT}^{ji}(t, \bk_2, \bk_1, \bv) \right |^2.
\label{eq:Fij}
\ee
The local small-scale power spectrum $P_B(\bk, \bv)$ defined in Eq.~\eqref{eq:PB} is not statistically isotropic, as it depends on the direction $\hk$ through $\bv\cdot\hk$. Note that, unlike the matter power spectrum, which depends on relative velocity only through its projection $\bv \cdot \hk$ along $\hk$ \cite{Tseliakhovich:2010bj}, the local small-scale power spectrum of the magnetic field depends on the relative magnitude $v_{\rm bc}$ and on its angle with $\hk$ \emph{separately}, i.e.~$P_B(\bk, \bv) = P_B(k, v_{\rm bc}, \hat{v}_{\rm bc} \cdot \hk)$.

A more physically meaningful quantity is the direction-averaged power spectrum, which by slight abuse of notation we denote $P(k, v_{bc})$, defined as
\barr
P_B(k, v_{bc}) &\equiv& \int \frac{d^2 \hk}{4 \pi} P_B(\bk, \bv) \nonumber\\
&=& \frac12 \int_{-1}^1 d (\hat{v}_{\rm bc} \cdot \hk) P_B(k, v_{\rm bc}, \hat{v}_{\rm bc} \cdot \hk). \label{eq:PB-vbc}
\earr
The isotropic magnetic field power spectrum only depends on the magnitude of the local relative velocity. Lastly, the global magnetic field power spectrum $P_B(k)$ is obtained by averaging $P_B(k, v_{\rm bc})$ over the Gaussian distribution of relative velocities at $z = 1060$, with variance per axis \cite{Tseliakhovich:2010bj} $\sigma_{\rm 1d}^2 = \langle v_{\rm bc}^2 \rangle/3 \approx (17 ~\textrm{km/s})^2$, explicitly, 
\be
P_B (k) = \sqrt{\frac{2}{\pi \sigma_{\rm 1d}^3}} \int_0^{\infty}  d v_{\rm bc} ~v_{\rm bc}^2~ e^{- v_{\rm bc}^2/2 \sigma_{\rm 1d}^2} P_B(k, v_{\rm bc}). \label{eq:PB-avg}
\ee
From the magnetic field power spectrum, one may obtain the overall variance of the magnetic field,
\be
\langle B^2 \rangle = \int \frac{d^3 k}{(2 \pi)^3} P_B(k) = \int d\ln k~ \frac{k^3}{2 \pi^2} P_B(k).   
\ee
Whenever we quote our results for $\langle B^2 \rangle$, we take an upper cutoff $k_{\max} = 10^4$ Mpc$^{-1}$ in this integral. 

We see that obtaining the magnetic field power spectrum at a given redshift and wavenumber requires computing a 5-dimensional integral: three angular integrals (corresponding to the relative angles between $\hk, \hk_1$ and $\hat{v}_{\rm bc}$), one integral over the magnitude of $k_1$, and one over the distribution of $v_{\rm bc}$. Moreover, the integrand in Eq.~\eqref{eq:PB} itself depends on time (or redshift) integrals of products of transfer functions. This therefore represents a very significant computational challenge, which requires careful optimization to remain tractable. We provide the details of our numerical implementation in Appendix \ref{app:numerics}.

\subsection{Application to nearly scale-invariant adiabatic initial conditions}
\label{sec:PB-ad}

In this section, we show the power spectrum of magnetic field produced by perturbations in electron density and gas temperature seeded by ``standard" primordial adiabatic perturbations, with a nearly scale-invariant primordial power spectrum
\be
\frac{k^3}{2 \pi^2} P^{\rm AD, std}(k) = A_s (k/k_p)^{n_s -1}, \label{eq:Pad-std}
\ee
where the pivot scale is $k_p = 0.05$ Mpc$^{-1}$, the amplitude is $A_s = 2.1 \times 10^{-9}$, and the spectral index is $n_s = 0.9665$, which are the Planck CMB-anisotropy best-fit values.

\begin{figure}[t!]
\centering
\includegraphics[width=\linewidth, trim= 10 20 10 10]{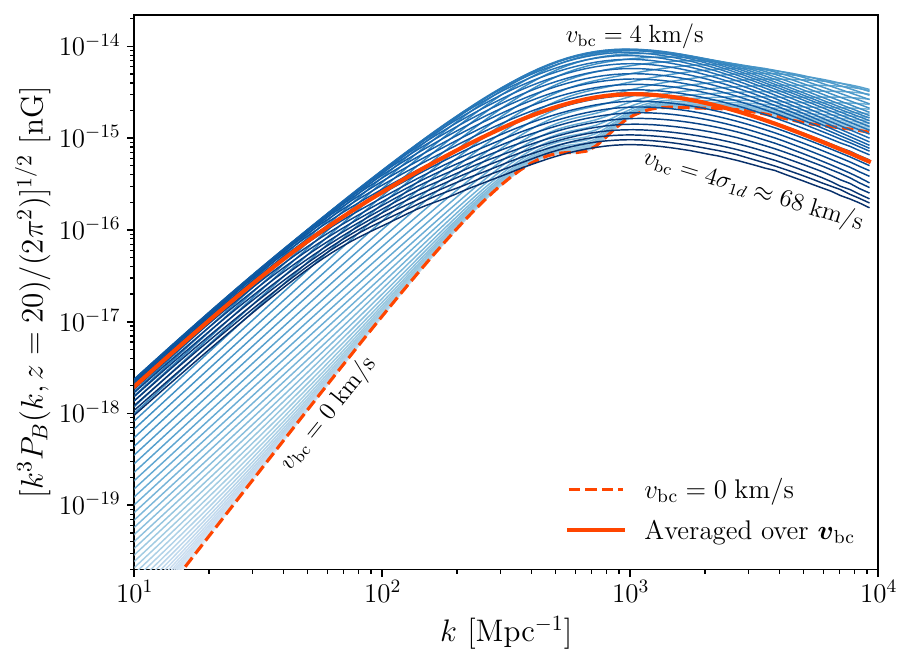}
\caption{Power spectrum of the magnetic field produced by the Biermann-battery mechanism with primordial adiabatic perturbations, extrapolating the Planck CMB best-fit cosmology \cite{Planck2018} to small scales, at $z=20$. The gradually colored blue curves correspond to the local isotropic power spectra $P_B(k, v_{\rm bc})$ defined in Eq.~\eqref{eq:PB-vbc}, for relative velocities $v_{\rm bc}\in [0, 4\sigma_{1d}]$, where $\sigma_{1d}\approx17\;\text{km/s}$. The color of $P_B(k, v_{\rm bc})$ with $v_{\rm bc}=0$ starts with light blue and gets darker as $v_{bc}$ increases being dark blue with $v_{\rm bc}=4\sigma_{1d}$. Overall, the amplitude of $P_B(k, v_{\rm bc})$ increases rapidly at first with small $v_{\rm bc}$'s, and after it peaks with $v_{\rm bc}\approx 4\;\text{km/s}$ at $k\sim10^3\;\text{Mpc}^{-1}$ the amplitude decreases with $v_{\rm bc}$. The solid orange line show the $v_{\rm bc}$-averaged result $P_B(k)$ defined in Eq.~\eqref{eq:PB-avg}. For reference, the dashed orange line shows $P_B(k, v_{\rm bc} = 0)$, the power spectrum one would obtain if neglecting relative velocities.
}
\label{fig:PB-ad-20}
\end{figure}

We present our result in Fig.~\ref{fig:PB-ad-20}, where we show the rms magnetic field per logarithmic scale, $\sqrt{k^3 P_B(k)/2\pi^2}$, at $z=20$. We see that it peaks around comoving wavenumber $k \sim 10^3\;\text{Mpc}^{-1}$, with an overall rms $\langle \bm{B}^2\rangle^{1/2} \simeq 4\times10^{-15}$nG. Accounting for relative velocities does not significantly shift the peak of $P_B$, but significantly enhances power on scales $k \lesssim 10^3$ Mpc$^{-1}$. In particular, non-vanishing relative velocities change the large-scale (small-$k$) asymptotic scaling of $\sqrt{k^3 P_B(k)}$ from $\propto k^{7/2}$ to $\propto k^{5/2}$. 

These asymptotic behaviors can be understood as follows. First, let us note that the reality condition of real-space perturbations implies that $\mathcal{T}_\alpha(k, - {\bv \cdot \hk}) = \mathcal{T}_\alpha^*(k, {\bv \cdot \hk})$ for any perturbation $\alpha$. Therefore, $\mathcal{F}_{eT}^{\rm AD, AD}(\bk_1, -\bk_1) = \Big{|}2~\textrm{Im}\left(X^{\rm AD, AD}_{eT}(\bk_1, - \bk_1)\right)\Big{|}^2$.

For $v_{\rm bc} = 0$, all transfer functions are real, and as a consequence, $\mathcal{F}_{eT}^{\rm AD, AD}(\bk_1, -\bk_1, \bv = 0) = 0$, implying that, at lowest order in $k$, $\mathcal{F}_{eT}^{\rm AD, AD}(\bk_1, \bk-\bk_1, \bv = 0) \propto k^2$ for $k \ll k_1$. From Eq.~\eqref{eq:PB}, we thus find $P_B(k) \propto k^4$ for $k$ much smaller than the peak of the integrand (which is around the Jeans scale $k_J$ for adiabatic initial conditions), explaining the observed scaling $\sqrt{k^3 P_B(k)} \propto k^{7/2}$. For $v_{\rm bc} \neq 0$, transfer functions acquire a non-vanishing imaginary part (see Fig.~\ref{fig:transfer}). This implies that $\mathcal{F}_{eT}^{\rm AD, AD}(\bk_1, -\bk_1, \bv \neq 0) \neq 0$. As a consequence, Eq.~\eqref{eq:PB} gives $P_B(k) \propto k^2$ for $k \ll k_J$, implying $\sqrt{k^3 P_B(k)} \propto k^{5/2}$, as observed in Fig.~\ref{fig:PB-ad-20}.

\begin{figure}[t!]
\centering
\includegraphics[width=\linewidth, trim= 10 20 10 10]{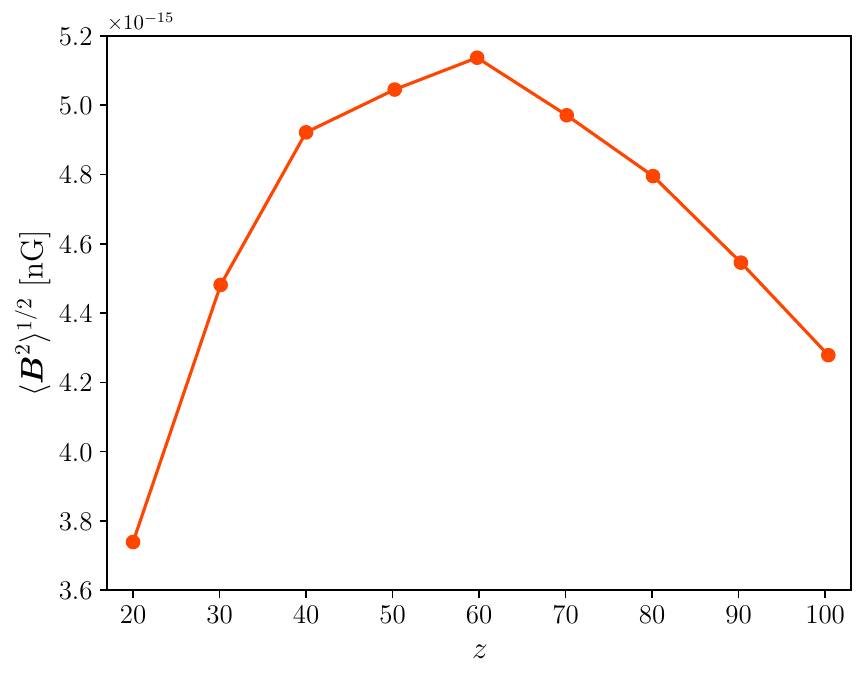}
\caption{The rms of velocity-averaged magnetic field produced by the Biermann-battery mechanism with primordial adiabatic perturbations, extrapolating the Planck CMB best-fit cosmology \cite{Planck2018} to small scales, as a function of redshift. The rms at $z=20$ is that of the velocity-averaged magnetic field shown as an orange line in Fig.~\ref{fig:PB-ad-20}. Note that the redshift-evolution of the rms is determined by both the production rate of magnetic field and the dilution of the existing field due to the expansion of the Universe.}
\label{fig:B-ad-z}
\end{figure}

Let us recall that our results should only be trusted for $k \lesssim 10^3$ Mpc$^{-1}$, since for smaller scales finite photon propagation effects may significantly alter the evolution of perturbed recombination \cite{Venumadhav_2015}.

The redshift evolution of the rms magnetic field is determined by a competition between the Biermann-battery production rate and dilution by cosmological expansion. We show this evolution in Fig.~\ref{fig:B-ad-z}, where we see that the rms magnetic field increases until $z \approx 60$, after which it starts decreasing. This interesting result could not easily be predicted analytically.

Our results are significantly different from those obtained in the pioneering work of NN13. Specifically, we obtain an overall magnetic field rms larger than theirs by at least one order of magnitude, and significantly different large-scale asymptotic behaviors for $P_B(k)$. We were able to identify some assumptions that may explain part of these differences. First, NN13 assume that the perturbations in free electron \emph{density} are equal to the perturbations in free-electron \emph{fraction}, i.e.~that $\delta_{n_e} = \delta_{x_e}$, instead of $\delta_{n_e} = \delta_{x_e} + \delta_b$. However, we find that, for adiabatic perturbations, $|\delta_{x_e}| \lesssim 0.2 |\delta_b|$ at $z \lesssim 50$ on all scales $k \geq 1$ Mpc$^{-1}$. Hence, NN13's assumption significantly underestimates the free-electron density perturbation, which may explain why they obtain a lower magnetic field amplitude than we do. Second, NN13 estimate the angle-average power spectrum $P_B(k, v_{\rm bc})$ by substituting\footnote{Private communication from S.~Naoz.} $\mathcal{T}_\alpha(\bk, \bv) \rightarrow \langle |\mathcal{T}_\alpha(\bk, \bv)|^2 \rangle^{1/2}$ in Eq.~\eqref{eq:X}, where the average is over the angle between $\bk$ and $\bv$. This does not account for correlations between transfer functions at different wavenumbers for a given relative velocity, as is correctly accounted for when we average $P_B$ itself (which is \emph{quartic} in transfer functions) over angles. This assumption of NN13 explains why they obtain the same large-scale asymptotic scaling for $P_B(k)$ regardless of the value of $v_{\rm bc}$: their averaging procedure spuriously eliminates the imaginary part of $X^{\rm AD, AD}_{eT}$, which, as we show above, is responsible for the different asymptotic scaling of $P_B(k, v_{\rm bc} \neq 0)$. Neither one of these assumptions (nor the minor differences in our evolution equations for $x_e$) explain the asymptotic behavior of $P_B(k)$ found by NN13, however: their Fig.~2 shows $\sqrt{k^3 P_B(k)} \propto k^{1.7}$ on large scales, implying $P_B(k) \propto k^{0.4}$, independently of $v_{bc}$. This result is inconsistent with the fact that $P_B(k)$ should scale \emph{at least} quadratically in $k$ on large scales, as can be seen e.g.~from Eq.~\eqref{eq:PB}. Our new results should thus be seen as updating and superseding those of NN13.

\section{Magnetic field as a probe of small-scale initial conditions}
\label{sec:PB-probe}

\subsection{Setup}

We now compute the magnetic field generated by the superposition of the ``standard" (CMB-extrapolated) adiabatic initial conditions with power spectrum given in Eq.~\eqref{eq:Pad-std}, with a non-standard power spectrum for a single type of primordial perturbation $i$ (possibly adiabatic as well) with a Dirac-delta peak at scale $k_0$:
\barr
P^i(k) = \frac{2 \pi^2}{k_0^2} \Delta^2(k_0) \delta_{\rm D}(k - k_0). \label{eq:spike}
\earr
The resulting magnetic field power spectrum is then 
\barr
P_B(\bk, \bv) &=& P_B^{\rm std}(\bk, \bv) \nonumber\\
&+& P_B^{i \times \rm std}(\bk, \bv) + P_B^{i, i}(\bk, \bv),~~ 
\label{eq:total-PB-dirac}
\earr
which includes the cross term, linear in $\Delta^2(k_0)$,
\barr
P_B^{i \times \rm std }(\bk, \bv) \equiv  \frac{c^2}{a^4 e^2} k_0^2 \Delta^2(k_0) \int \frac{d^2 \hk_1}{4 \pi} |\hk_1 \times \bk|^2 \nonumber\\
 \times P^{\rm AD, std}(|\bk - k_0 \hk_1|)\mathcal{F}_{eT}^{i, \rm AD}(k_0 \hk_1, \bk - k_0 \hk_1, \bv),
\earr
and a term quadratic in $\Delta^2(k_0)$, given by
\barr
P_B^{i, i}(\bk, \bv) \equiv \frac{\pi}{4} \left[\frac{c}{a^2 e} \Delta^2(k_0)\right]^2 k \left[1 - (k/2 k_0)^2\right] \nonumber\\
\times \int_0^{2 \pi} d \phi_1 ~\mathcal{F}_{eT}^{i,i}(k_0 \hk_1, \bk - k_0 \hk_1, \bv)\big{|}_{\hk_1 \cdot \hk = k/2k_0}, 
\earr
for $k < 2 k_0$, and vanishing otherwise, where $\phi_1$ is the azimuthal angle of $\hk_1$ in the spherical polar coordinate system where $\hk$ is the zenith, and the plane spanned by $(\hk, \hat{v}_{\rm bc})$ has $\phi_1 = 0$. The term $P_B^{i, i}(k)$ peaks near $k \sim k_0$.

In what follows we apply these results to spikes in the small-scale adiabatic or isocurvature perturbations, saturating existing upper limits, to show the maximum magnetic field they might generate. 

\subsection{Application to enhanced small-scale adiabatic perturbations}
CMB anisotropy data only probes relatively large-scale ($k \lesssim 0.3\;\text{Mpc}^{-1}$) adiabatic modes. On smaller scales, the most stringent constraints to date can be derived from upper bounds to CMB spectra distortions. Ref.~\cite{Chluba:2012we} provide simple analytic expressions for the chemical potential $\mu$ and Compton-$y$ distortion sourced by small-scale adiabatic perturbations with a Dirac-Delta spike as in Eq.~\eqref{eq:spike}, accurate to $20\%$ for $k \gtrsim 5$ Mpc$^{-1}$:
\barr
\mu &\approx& 2.2~\left(e^{- k_0/k_\mu} - e^{- (k_0/k_y)^2}\right) \Delta^2(k_0), \\
y &\approx& 0.4 ~e^{-(k_0/k_y)^2} \Delta^2(k_0), 
\\
k_\mu &=& 5400~\textrm{Mpc}^{-1}, \ \ \ k_y = 31.6 ~\textrm{Mpc}^{-1}.
\earr
COBE-FIRAS \cite{Fixsen_96} constrains $|\mu| \leq 9 \times 10^{-5}$ and $|y| \leq 1.5 \times 10^{-5}$ at the 95\% confidence interval. This implies that, the bound on $\Delta^2(k_0)$ is approximately
\barr
\Delta^2_{\rm AD}(k_0) \lesssim 4 \times 10^{-5} ~e^{k_0/k_\mu}.
\label{eq:ad-SD-limit}
\earr
This expression recovers the bounds from either $\mu$ or $y$ limits in the region where either one dominates.

\begin{figure}[t!]
\centering
\includegraphics[width=\linewidth, trim= 10 20 10 10]{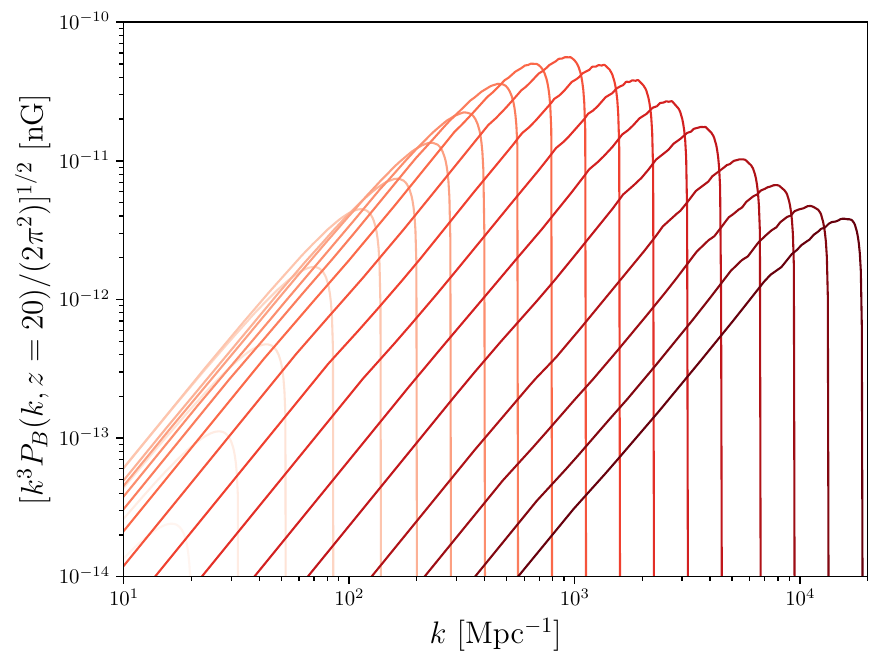}
\caption{Velocity-averaged magnetic field power spectrum at $z=20$ produced by adiabatic primordial perturbations with a Dirac-Delta spike at $k=k_0\in [10,10^4]\;\text{Mpc}^{-1}$ whose amplitude saturates CMB spectral distortion upper limits, Eq.~\eqref{eq:ad-SD-limit}. The color of the curves starts with light red at $k_0=10\;\text{Mpc}^{-1}$ and gets darker as $k_0$ increases. Each curve peaks near $k \sim k_0$.}
\label{fig:PB-ad-20-dirac}
\end{figure}

Fig.~\ref{fig:PB-ad-20-dirac} shows the velocity-averaged $B$-field power spectrum at $z=20$ produced by adiabatic primordial perturbations with a Dirac-Delta spike saturating spectral distortion upper limits (in addition to the ``standard" nearly scale-invariant power spectrum), for varying spike positions $k_0$. We find that magnetic fields can be as large as $\sim 5 \times 10^{-11}$ nG, four orders of magnitude larger than that generated by standard adiabatic initial conditions, without violating CMB spectral-distortion constraints. This maximum rms is attained for $k_0 \sim 10^3$ Mpc$^{-1}$.

\subsection{Application to small-scale isocurvature perturbations}

We now compute the magnetic field power spectrum generated by various small-scale isocurvature perturbations saturating current limits. For BI, BCI and CIPs, the most stringent upper limits come from Big-Bang Nucleosynthesis (BBN) constraints, first derived in Ref.~\cite{Inomata:2018htm} and revised in Ref.~\cite{Lee:2021bmn}, which limit the primordial baryon variance on scales larger than the neutron diffusion scale, $k \lesssim 4 \times 10^8$ Mpc$^{-1}$, implying 
\be
\Delta^2_i(k_0) < 0.019, ~~~~ i = \textrm{BI, BCI, CIP} ~~(95\% ~\textrm{c.l.})\label{eq:BBN}
\ee
For pure CDM isocurvature (CI), the BBN bound does not apply. Instead, the strongest constraint is the one we obtained in Ref.~\cite{Lee:2021bmn} based on the perturbation of the average recombination history, which is of order $\Delta_{\rm CI}^2(k_0) \lesssim 0.7$ for $k_0 \lesssim 500$ Mpc$^{-1}$, and rapidly degrades for smaller scales. Given that linear perturbation theory should not apply for such large amplitudes, the precise value of the bound is not robust. Therefore, for illustration purposes, we simply compute the magnetic field generated with $\Delta_{\rm CI}^2(k_0)=1$.

\begin{figure}[t!]
\centering
\includegraphics[width=\linewidth, trim= 10 20 10 10]{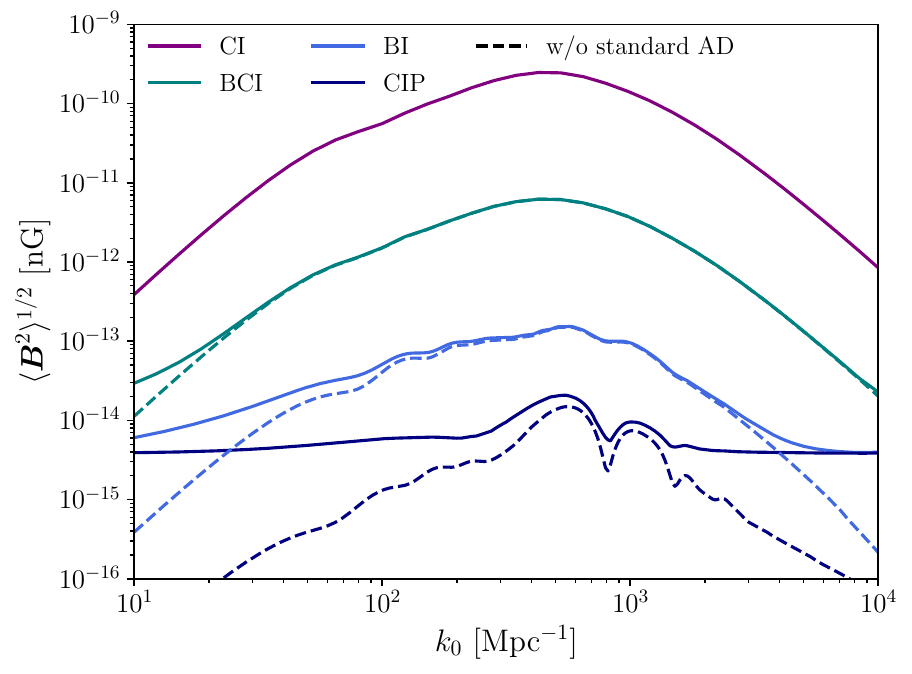}
\caption{The rms magnetic field at $z=20$ produced by the Biermann-battery mechanism with a Dirac-delta primordial isocurvature power spectrum peaked at $k=k_0$ [Eq.~\eqref{eq:spike}]. Note that while BCI, BI, and CIP are under BBN bound \cite{Inomata:2018htm,Lee:2021bmn} [Eq.~\eqref{eq:BBN}] as there is no robust constraint on the amplitude of CI, we set $\Delta^2(k_0)=1$ for CI. Dashed lines only include contribution purely from isocurvature perturbations, and solid lines include the contribution from ``standard" adiabatic perturbations. Note that the rms of $\bm{B}_{\rm std}$ is $\sim4\times 10^{-15}\;\text{nG}$.}
\label{fig:B-iso}
\end{figure}

We show in Fig.~\ref{fig:B-iso} the rms magnetic field $\langle \bm{B}^2 \rangle^{1/2}$ generated by small-scale isocurvature perturbations saturating the limits described above, as a function of the peak scale $k_0$. We find that, under current constraints, primordial isocurvature perturbations have the potential to enhance $\langle \bm{B}^2 \rangle^{1/2}$ up to $\sim 3 \times 10^{-12}\;\text{nG}$ for BCI, $\sim 10^{-13}$ nG for BI, and $\sim 10^{-14}$ nG for CIP. Since CI are not well constrained on these scales, they may source a magnetic field with rms up to $\sim 3 \times 10^{-10}$ nG. Note that as $P_B^{i,i}$ is the dominant contribution for \{BI, BCI, CI\}, the rms magnetic field scales as $\Delta_i^2(k_0)$ while $\langle \bm{B}^2\rangle^{1/2} \gg \langle \bm{B}_{\rm std}^2\rangle^{1/2}$, where $\bm{B}_{\rm std}$ is the magnetic field produced with ``standard" adiabatic initial conditions.

One would naively think that the magnetic field from isocurvature perturbations should be roughly a factor of $\Delta^2(k_0)/A_s$ larger than that sourced by standard adiabatic perturbations. However, this is not the case as the growth rate of perturbations is in general much slower for isocurvature perturbations than for adiabatic perturbations (see Fig.~\ref{fig:transfer}), since CDM perturbations are not boosted by the gravitational potential sourced by photon perturbations in the radiation-dominated era. 

Let us note that CIPs were first considered as a source of magnetic fields through the Biermann-battery mechanism in Ref.~\cite{Flitter:2023xql}, whose authors considered CIPs with power-law primordial power spectra. Our treatment is much more detailed in several respects, leading to up to one order of magnitude difference with the results of Ref.~\cite{Flitter:2023xql} with equal assumptions for the CIP primordial power spectrum. We discuss the differences between our work and theirs in Appendix.~\ref{appendix:PB-cip}.

\section{Conclusion}
\label{sec:discussion}

In this paper we make the most accurate calculation to date of the magnetic field produced by a variety of primordial perturbations through the Biermann-battery mechanism. To do so, we consistently solve for the evolution of small-scale perturbations in electron density and baryon gas temperature, including the effect of large-scale relative velocity between CDM and baryon. 

We first calculate the magnetic field produced solely from standard, nearly-scale invariant primordial adiabatic perturbations. Our results update and supersede those of the pioneering work NN13 \cite{Naoz:2013wla}, as we correctly compute the electron density perturbation, and account for the full angular dependence of the magnetic field power spectrum on relative velocities. We show that the resulting ``standard" magnetic field can reach an rms of a few times $10^{-15}$ nG at $z=20-100$, with maximal fluctuations on comoving scales $k \sim 10^3$ Mpc$^{-1}$. This magnetic field can serve as a seed field to be later amplified by dynamo processes, possibly explaining the magnetic fields present in galaxies and galaxy clusters today \cite{Durrer:2013pga}. 

In addition, using the fact that the main contribution to magnetic field comes from perturbations on very small scales $k \sim 10^2-10^3$ Mpc$^{-1}$, we argue that measuring magnetic fields can be a useful probe for initial conditions of the Universe on such small scales. We first consider enhanced small-scale primordial adiabatic perturbations, and show that they can source magnetic fields as large as $\sim 5 \times 10^{-11}$ nG without violating current CMB spectral distortion constraints. We similarly compute the magnetic field sourced by a variety of small-scale isocurvature perturbations (baryon, CDM, joint baryon and CDM, and compensated baryon-CDM isocurvature perturbations). We find that such non-standard perturbations could generate magnetic fields as large as $\sim 10^{-14} - 10^{-10}$ nG at cosmic dawn, depending on the specific type of isocurvature mode, without violating current BBN and perturbed recombination constraints. While these are extremely weak magnetic fields, below current detection capabilities, they may still be a useful observable to consider in the future, as they could provide a new window into the poorly constrained small-scale initial conditions of the Universe.

\section*{Acknowledgements}

We thank Marc Kamionkowski for useful conversations and encouragements, Smadar Naoz for providing details and data relevant to her work on the topic, and Jordan Flitter for providing his results for a comparison with our work, as well as sending detailed comments about this manuscript.  N.\;L. is supported by the Center for Cosmology and Particle Physics at New York University through the James Arthur Graduate Associate Fellowship. Y.\;A.-H. is a CIFAR-Azrieli Global Scholar and acknowledges support from the Canadian Institute for Advanced Research (CIFAR), and is grateful for being hosted by the USC departement of physics and astronomy while on sabbatical. This work was supported in part through the NYU IT High Performance Computing resources, services, and staff expertise.

\begin{appendix}

\section{Derivation of the Biermann-battery mechanism} \label{app:Biermann}

In this appendix we provide a pedagogical derivation of the Biermann-battery mechanism, in the cosmological context. For simplicity we do not account for cosmic expansion in this derivation (alternatively, it can be thought of as a derivation in a locally-inertial coordinate system), and restore dependence on scale factors in the final result. We work in units where $c = 1$.

\subsection{Electron-proton slip and current density}

We consider free electrons and free protons as two distinct ideal fluids (a more rigorous approach would use kinetic theory and follow their distribution functions). For each fluid $i = e, p$, we denote by $n_i$ its number density, $\rho_i = m_i n_i$ its mass density, $\bs{v}_i$ its velocity, $T_i$ its temperature and $P_i = n_i T_i$ its pressure. Assuming helium is fully neutral, all of these fields are nearly identical for electrons and protons, except for their mass densities, $\rho_p \approx (m_p/m_e) \rho_e \gg \rho_e$.

In the presence of an electric and magnetic field, the momentum equations satisfied by each fluid are 
\barr
\frac{d \bs{v}_e}{dt} &=& - \bs{\nabla} \phi - \frac{\bs{\nabla}P_e}{\rho_e} - \frac{e}{m_e} \left( \bs{E} + \bs{v}_e \times \bs{B}\right) \nonumber\\
&+& \Gamma_{ep} (\bs{v}_p - \bs{v}_e), \\
\frac{d \bs{v}_p}{dt} &=& - \bs{\nabla} \phi - \frac{\bs{\nabla}P_p}{\rho_p} + \frac{e}{m_p} \left( \bs{E} + \bs{v}_p \times \bs{B}\right)\nonumber\\
&-& \frac{\rho_e}{\rho_p} \Gamma_{ep} (\bs{v}_p - \bs{v}_e), 
\earr
where $e$ is the elementary charge, and we neglected photon drag. Aside from the obvious gravitational, pressure and electromagnetic forces, the last term, proportional to the relative velocity between electrons and protons, represents the drag force arising from frequent Coulomb interactions between the two species -- it is this term which, in practice, forces electrons and protons to be behave very nearly as a single fluid. Subtracting the two equations, and only keeping terms at lowest order in $m_e/m_p \ll 1$, 
we obtain an equation for the ``slip", i.e.~the velocity difference between the two species:
\barr
\frac{d}{dt}(\bs{v}_p - \bs{v}_e) + \Gamma_{ep} (\bs{v}_p - \bs{v}_e) \nonumber\\
= \frac{\bs{\nabla} P_e}{\rho_e} + \frac{e}{m_e} \left( \bs{E} + \bs{v}_e \times \bs{B}\right). \label{eq:slip}
\earr
Let us now estimate $\Gamma_{ep}$: it is of the order of the rate of interactions between electrons and protons, namely $\Gamma_{ep} \sim n_e \langle \sigma_{\rm C}(v_{\rm rel}) v_{\rm rel} \rangle$, where $\sigma_{\rm C}$ is the Coulomb interaction cross section and $v_{\rm rel}$ is the microscopic (in contrast with the macroscopic averages $\bs{v}_p, \bs{v}_e$) relative velocity of electrons and protons -- for completeness, exact expressions for the momentum-exchange rate can be found in Refs.~\cite{YAH_2019, Ali-Haimoud:2023pbi}. The Coulomb scattering cross section is approximately $\sigma_{\rm C}(v) \sim (e^2/(m_e v^2))^2 \sim \sigma_{\rm T}/v^4$, where $\sigma_{\rm T}$ is the Thomson cross section. Since microscopic relative velocities between electrons and protons are dominated by the electrons' thermal velocities, we find
\be
\Gamma_{ep} \sim n_e \sigma_{\rm T} (m_e/T_e)^{3/2}. \label{eq:Gamma_ep}
\ee
We may estimate the ratio of this rate relative to the Hubble rate by noting that the Thomson scattering rate $n_e \sigma_{\rm T}$ becomes comparable to the Hubble rate around redshift $z \sim 10^3$, at which point $x_e \sim 0.1$. Hence, we find (assuming $H(z) \propto z^{3/2}$ during matter domination, and with $n_e \propto x_e z^3$),
\be
\frac{\Gamma_{ep}}{H} \sim \frac{x_e}{0.1} (z/10^3)^{3/2}(m_e/T_e)^{3/2}.
\ee
The electron temperature is approximately 
\be
T_e \simeq \begin{cases} 0.25 ~\textrm{eV} \times (z/10^{3}),  \ \ \ \ z \gtrsim 120,\\
0.03 ~\textrm{eV} \times (z/120)^2,  \ \ \  z \lesssim 120,
\end{cases} \label{eq:Te}
\ee
where the transition at $z \approx 120$ corresponds to thermal decoupling from photons. Therefore, we find, at $z \lesssim 120$, 
\be
\frac{\Gamma_{ep}}{H} \sim  10^7~\frac{x_e}{3 \times 10^{-4}} (z/100)^{-3/2}, 
\ee
where we have normalized $x_e$ to its freeze-out abundance. We thus see that, at all relevant times, $\Gamma_{ep} \gg H$, implying that electrons and protons are tightly coupled. This allows us to estimate the slip relative velocity by solving Eq.~\eqref{eq:slip} in the quasi-steady-state approximation:
\be
\bs{v}_p - \bs{v}_e \approx \frac1{\Gamma_{ep}} \left[\frac{\bs{\nabla} P_e}{\rho_e} + \frac{e}{m_e} \left( \bs{E} + \bs{v}_e \times \bs{B}\right) \right].
\ee
This slip is an important quantity as it allows us to determine the electric current density $\bs{j} = e (n_p \bs{v}_p - n_e \bs{v}_e) \approx e n_e (\bs{v}_p - \bs{v}_e)$. We thus have found
\be
\bs{E} + \bs{v}_e \times \bs{B} + \frac{\bs{\nabla}P_e}{e n_e} \approx \Gamma_{ep} \frac{m_e}{e^2 n_e} \bs{j}. \label{eq:E-j}
\ee

\subsection{Combining with Maxwell's equations}
We now combine Eq.~\eqref{eq:E-j} with Maxwell's equations. Given a current density $\bs{j}$, the magnetic field satisfies the forced wave equation
\be
\bs{\nabla} \times \bs{j} = \frac1{4 \pi} \left[\partial_t^2 \bs{B} - \nabla^2 \bs{B}\right]. \label{eq:B-wave}
\ee
When considering scales much smaller than the Hubble radius, one may neglect the first term relative to the second one in the right-hand-side of Eq.~\eqref{eq:B-wave}. Taking the curl of Eq.~\eqref{eq:E-j} and using Maxwell-Faraday equation (and neglecting small fluctuations of the prefactor in front of $\bs{j}$), we obtain
\barr
- \partial_t \bs{B} + \bs{\nabla} \times (\bs{v}_e \times \bs{B}) + \frac1{e} \bs{\nabla} \times (\bs{\nabla} P_e/n_e) \nonumber\\
\approx -\Gamma_{ep} \frac{m_e}{e^2 n_e} \frac1{4\pi} \nabla^2 \bs{B} \equiv - \lambda_{ep} \nabla^2 \bs{B}, \label{eq:B-prelim}
\earr
where we have defined a characteristic lengthscale
\be
\lambda_{ep} \equiv \Gamma_{ep} \frac{m_e}{4 \pi e^2 n_e} \sim (e^2/m_e) (m_e/T_e)^{3/2},
\ee
where we used Eq.~\eqref{eq:Gamma_ep} and $\sigma_{\rm T} = (8\pi/3) (e^2/m_e)^2$. Inserting Eq.~\eqref{eq:Te}, we thus obtain, for $z \lesssim 120$
\be
\lambda_{ep} \sim 0.03~ \textrm{cm} ~ (z/100)^{-3}.
\ee
For fluctuations on a comoving scale $k$, the ratio of the term proportional to $\nabla^2 \bs{B}$ to the first term in Eq.~\eqref{eq:B-wave} is of order 
\barr
\frac{\lambda_{ep} |\nabla^2 \bs{B}|}{|\partial_t \bs{B}|} &\sim& \frac{\lambda_{ep} k^2}{a^2 H} \nonumber\\
&\sim& \left(\frac{k}{3 \times 10^{13} ~\textrm{Mpc}^{-1}}\right)^2 (z/100)^{-5/2}.~~~~~
\earr
Therefore, we see that, for any cosmological scales of interest, the term proportional to $\nabla^2 \bs{B}$ may be safely neglected in Eq.~\eqref{eq:B-prelim}. If we restrict ourselves to small perturbations in the electron density, we may moreover neglect the term $\bs{\nabla} \times (\bs{v}_e \times \bs{B})$. We have thus found
\be
\partial_t \bs{B} \approx \frac1{e} \bs{\nabla} \times (\bs{\nabla} P_e/n_e) = - \frac1{e} \frac{\bs{\nabla} n_e}{n_e^2} \times \bs{\nabla} P_e.
\ee
With $P_e = n_e T_e$, and $n_e = \ov{n}_e (1 + \delta_{n_e})$, $T_e = \ov{T}_{\rm gas}(1 + \delta_{T_{\rm gas}})$, we find
\be
\partial_t \bs{B} \approx - \frac{\ov{T}_{\rm gas}}{e} \bs{\nabla} \delta_e \times \bs{\nabla} \delta_{T_{\rm gas}}.
\ee
Re-expressing proper gradients in terms of comoving gradients, and accounting for the dilution of magnetic fields by $1/a^2$ due to cosmological expansion, we recover Eq.~\eqref{eq:dBdt-real}.

\section{Numerical implementation details} \label{app:numerics}

In this appendix we detail our numerical implementation of the magnetic field power spectrum.\\

$\bullet$ We first store the transfer functions of electron density perturbations $\delta_{n_e}$ and baryon gas temperature perturbations $\delta_{T_{\rm gas}}$ in tables. We note that, for large relative velocities or small scales, all perturbations are proportional to the rapidly-oscillating component $e^{i \phi(z, \bk\cdot \bv^{\rm ini})}$, with phase
\barr
\phi(z, \bk\cdot \bv^{\rm ini}) &\equiv& \int_{t_{\rm ini}}^t \frac{dt'}{a'}  \bk \cdot \bv(t') \nonumber\\
&=& (\bk \cdot \bv^{\rm ini}) \int_{z}^{z_{\rm ini}} \frac{dz'}{H(z')} \frac{1 + z'}{1 + z_{\rm ini}},
\earr
where $z_{\rm ini} \equiv 1060$. Instead of tabulating the rapidly-oscillating small-scale transfer functions themselves, we thus tabulate their slow-varying parts 
\be
\widetilde{\mathcal{T}}_\alpha^i(z, k, \bv^{\rm init} \cdot \hk) \equiv e^{- i \phi(z, \bk\cdot \bv^{\rm init})} \mathcal{T}_\alpha^i(z, k, \bv^{\rm init} \cdot \hk). \label{eq:slow-part} 
\ee
This allows us to get accurate interpolation results with a coarser table. For small-scale modes, we store the slow-varying parts $\widetilde{\mathcal{T}}$ for 1,300 wavenumbers $k\in[1,2\times10^4]\;\text{Mpc}^{-1}$, 200 values of $\bm{v}_{bc}\cdot\hat{k} \in [0,4\sigma_{1d}]$, and 300 redshift points in $z\in[20,z_{\rm ini}]$ logarithmically spaced in scale factor (or equivalently in $1+z$). For large-scale modes, we store transfer functions only with $v_{bc}=0$ for 450 wavenumbers $k\in[3\times10^{-4},1]\;\text{Mpc}^{-1}$ and 1,000 redshifts $z\in[20,z_{\rm ini}]$. When needed, we interpolate these tables using \textsc{Python} \textsc{Scipy} cubic spline interpolator.

$\bullet$ We compute the time integral in $X_{eT}^{ij}$ [Eq.~\eqref{eq:X}], using the trapezoidal rule with the 300 logarithmically-spaced redshifts in $z\in[20,z_{\rm ini}]$ at which we store the transfer functions at high $k$'s. This avoids having to do 3-dimensional interpolation: we only interpolate over $k$ and $\bm{v}_{bc}\cdot\hat{k}$ for each given redshift grid-point. In principle, this time integral should be extended to $t \rightarrow 0$ (or $z \rightarrow \infty$), but we checked that it is in fact already well converged when only including $z < z_{\rm ini}=1060$, since the integral is dominated by late times. 

$\bullet$ For given values of $v_{\rm bc}$ and $\hat{v}_{\rm bc}\cdot\hat{k}$, we compute the 3-dimensional $\bs{k}_1$-integral of Eq.~\eqref{eq:PB} as follows. We define the $z$-axis to be aligned with the vector $\bm{k}$, and let $\bv$ be on $x$-$z$ plane without loss of generality. We then perform the integral in spherical polar coordinates $k_1, \theta, \phi$, noting that the symmetry of the problem implies $\int_0^{2 \pi} d \phi = 2 \int_0^{\pi} d \phi$. We then perform 21-point Gaussian-Quadrature integration for $\phi \in [0,\pi]$. We rewrite the polar angle integral as $\int_{-1}^1d\cos\theta = -\int_0^{\sqrt{2}} 2x dx$, where we define $x\equiv \sqrt{1-\cos\theta}$, and then perform 21-point Gaussian Quadrature integration for $x\in[0,\sqrt{2}]$. This is to better handle a divergence which happens because the integrand scales as $(1-\cos\theta)^{-1/2}$ when $k_1=k$. Given a set of two angular variables, we integrate over $k_1$ from $k_{\rm min}=3\times10^{-4}\;\text{Mpc}^{-1}$ to $k_{\rm max}=10^4\;\text{Mpc}^{-1}$ using the 1-dimensional integrator in \textsc{Python} \textsc{Scipy} package (\texttt{scipy.integration.quad}). Finally, we compute the 2-dimensional angular integral by trapezoidal rule over sampled points of $\cos\theta$ and $\phi$.

$\bullet$ Finally, we compute the 2-dimensional integral over $v_{\rm bc}$ and $\hat{v}_{\rm bc}\cdot\hat{k}$, given in Eq.~\eqref{eq:PB-vbc} and Eq.~\eqref{eq:PB-avg}, by doing 11-point Gaussian Quadrature integration for each 1-dimensional integration.

\section{Comparison with Flitter et al.~2023 \cite{Flitter:2023xql}}
\label{appendix:PB-cip}

In this appendix, we review the assumptions taken in Ref.~\cite{Flitter:2023xql} (hereafter F+23) to compute the magnetic field generated by CIPs, and contrast them with our detailed calculation. 

On scales much larger than the baryon Jeans scale, baryon and CDM perturbations are known to remain constant in time for CIP initial conditions. Motivated by this result, F+23 assume a strictly constant baryon perturbation $\delta_b^{\rm CIP} = \Delta$ for all scales $k \le 400$ Mpc$^{-1}$, and compute the resulting late-time free-electron density perturbation, finding $\delta_e^{\rm CIP} \approx - 0.06 \Delta$. In contrast, we compute the full time evolution of the free-electron density perturbation, as a function of scale, and find that it significantly differs from (and can be significantly larger than) this constant value even at scales $k \ll 400$ Mpc$^{-1}$, as can be seen in Fig.~\ref{fig:transfer}.

In addition, F+23 neglect the gas temperature fluctuations sourced by CIP initial conditions, hence only account for the cross term $\delta_e^{\rm CIP} \delta_{T_{\rm gas}}^{\rm AD}$ in the magnetic field source. This assumption is self-consistent with their assumption of constant $\delta_b^{\rm CIP}$, implying $\theta_b^{\rm CIP} = 0$, which thus cancels the main source of gas temperature fluctuations in Eq.~\eqref{eq:evolution}. In contrast, we self-consistently account for CIP-sourced gas temperature fluctuations, which are in fact comparable to free-electron density perturbations even for scales significantly larger than the Jeans scale (see Fig.~\ref{fig:transfer}). We thus effectively account for two more terms in the magnetic field source: $\delta_e^{\rm AD} \delta_{T_{\rm gas}}^{\rm CIP}$ and $\delta_e^{\rm CIP} \delta_{T_{\rm gas}}^{\rm CIP}$.

Moreover, F+23 assume that the AD-sourced gas temperature fluctuations strictly follow the baryon density fluctuations, $\delta_{T_{\rm gas}}^{\rm AD} = \frac23 \delta_b^{\rm AD}$. In turn, baryon density fluctuations are assumed to precisely match the total matter density fluctuation up to a constant Jeans scale $k_J = 200$ Mpc$^{-1}$, and to be suppressed by a factor $(k_J/k)^2$ for smaller scales \cite{Naoz:2005pd,Naoz:2010hg}, i.e.~modeled as $\delta_b^{\rm AD}(k) = \min[1, (k_J/k)^2] \times \delta_m^{\rm AD}(k)$. F+23 then model the matter power spectrum using \textsc{class} at large scales, extended to small scales ($k>100\;\text{Mpc}^{-1}$) by the Bardeen-Bond-Kaiser-Szalay fitting function \cite{Bardeen:1985tr} with baryon corrections of Ref.~\cite{Sugiyama:1994ed}. In contrast, we self-consistently compute the electron density and gas temperature perturbations for AD initial conditions, without imposing a constant bias with respect to density fluctuations nor any approximate cutoff at the Jeans scale.

Last but not least, F+23 do not account for relative velocities between CDM and baryons. Interestingly, these supersonic relative motions break the exact cancellation of the gravitational potential, thus allowing density perturbations to grow on scales larger than the baryon Jeans scale. This effect is thus qualitatively the opposite from what happens for adiabatic perturbations \cite{Tseliakhovich:2010bj}. However, it does not necessarily lead to an overall increase of the magnetic field power, since it pushes gas temperature fluctuations closer to the adiabatic limit $\delta_{T_{\rm gas}} \rightarrow \frac23 \delta_b$, and makes the ratio $\delta_{T_{\rm gas}}/\delta_e$ closer to being scale-invariant, i.e.~``aligns" more the gradients of gas temperature and electron density fluctuations.

As a concrete example, we consider a scale-invariant primordial power spectrum for CIPs, $P_{\rm CIP}(k) = \frac{2 \pi^2}{k^3} A_{\rm CIP}$, corresponding to the spectral index $\alpha=-3$ case in F+23. In that case the upper limit on $A_{\rm CIP}$ is dominated by large-scale CMB-anisotropy constraint, which with our convention\footnote{Ref.~\cite{Smith:2017ndr} define $P_{\rm CIP}(k) = A_{\rm CIP}/k^3$ instead of our $2 \pi^2 A_{\rm CIP}/k^3$} is $A_{\rm CIP} \leq 0.88  \times 10^{-3}$ \cite{Smith:2017ndr}.

\begin{figure}[t!]
\centering
\includegraphics[width=\linewidth, trim= 10 20 10 10]{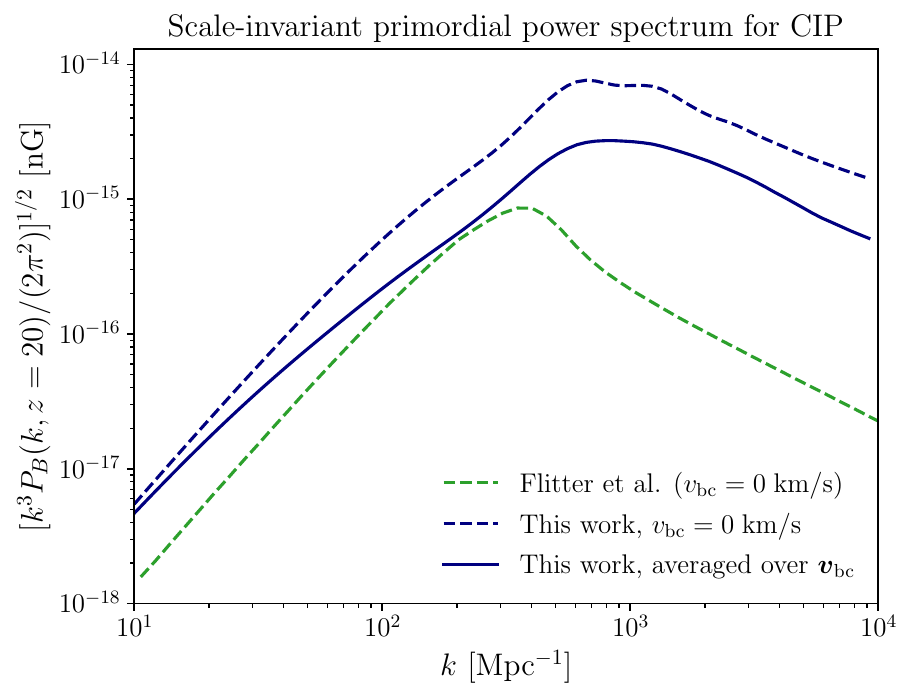}
\caption{Comparison of our results with those of F+23, assuming a scale-invariant primordial power spectrum for CIPs (the case of $\alpha=-3$ in their definition). The green dashed line is the reproduction of calculation done in F+23, and navy curves are from our calculations, only including the contributions from CIPs (and not the contribution from standard adiabatic fluctuations alone).}
\label{fig:PB-cip}
\end{figure}

We see in Fig.~\ref{fig:PB-cip} that the magnetic field from our calculations without any approximations is an order of magnitude larger than the result of F+23 at $z=20$. Note that the difference would have even been larger had we not included relative velocities, which lower the rms magnetic field by a factor of $\sim 2\text{--}3$. Also note that, while the observable magnetic field also includes contributions from standard adiabatic modes alone [shown to be $\mathcal{O}(10^{-15}\;\text{nG})$ in Fig.~\ref{fig:PB-ad-20}], for a fair comparison, in Fig.~\ref{fig:PB-cip} we only include contributions from CIPs, sourced by $\delta_e^{\rm AD}\delta_{T_{\rm gas}}^{\rm CIP}$, $\delta_e^{\rm CIP}\delta_{T_{\rm gas}}^{\rm AD}$, and $\delta_e^{\rm CIP}\delta_{T_{\rm gas}}^{\rm CIP}$. 

\end{appendix}

\bibliography{mybib}

\end{document}